\documentclass[twocolumn,showpacs,preprintnumbers,amsmath,amssymb]{revtex4}
\usepackage{graphicx}% Include figure files
\usepackage{dcolumn}% Align table columns on decimal point
\usepackage{bm}% bold math
\begin{document}

\title{Feed-forward control for quantum state protection against decoherence}

\author{Chao-Quan Wang$^{1}$}%
\author{Bao-Ming Xu$^{1}$}
\author{Jian Zou$^{1}$}%
\email{zoujian@bit.edu.cn}
\author{Zhi He$^{2}$}%
 \author{Yan Yan$^{1}$}
 \author{Jun-Gang Li$^{1}$}%
\author{Bin Shao$^{1}$}%
 \affiliation{$^{1}$School of Physics, Beijing Institute of Technology, Beijing 100081, China \\ $^{2}$College of Physics and
Electronics, Hunan University of Arts and Science, Changde 415000,
China}%
 \date{\today}
\begin{abstract}
We propose a novel scheme of feed-forward control and its reversal for protecting quantum state against decoherence. Before the noise channel our pre-weak measurement and feed-forward are just to change the protected state into the state almost immune to the noise channel, and after the channel our reversed operations and post-weak measurements are just to restore the protected state. Unlike most previous state protection schemes, ours only concerns the noise channel and does not care about the protected state. We show that our scheme can effectively protect unknown states, nonorthogonal states and entangled states against amplitude damping noise. Our scheme has dramatic merits of protecting quantum states against heavy amplitude damping noise, and can perfectly protect some specific nonorthogonal states in an almost deterministic way, which might be found some applications in current quantum communication technology. And it is most important that our scheme is experimentally available with current technology.
\end{abstract}

\maketitle

\section{\label{sec:level1}INTRODUCTION\protect\\}
The dynamics of open quantum systems have been extensively studied.
One of the major tasks is how to avoid the effect of decoherence due
to interactions between system and its environment. To reduce the
effect of decoherence, a large number of strategies have been
proposed including decoherence-free subspaces \cite{1,2,3}, quantum
error correction code \cite{4,5,6}, dynamical decoupling
\cite{7,8,9}, quantum feedback control \cite{10,11,12,13,14,15} etc.
Controlling the dynamics of quantum system is the central goal in
many areas of quantum physics. Recent experimental advances have
enabled individual system to be monitored and manipulated at quantum
level in real time \cite{16,17,18,19,20}, which makes quantum
control more and more practical and realizable. Among different
quantum control schemes, quantum feedback control is widely studied
to improve the stability and robustness of the system, which is
based on feeding back the measurement results to alter the future
dynamics of quantum systems. The general framework of quantum
feedback control was introduced by Wiseman and Milburn \cite{10,11},
leading to relevant experimental achievements \cite{21,22,23}. But
information gain and disturbance in quantum systems are always
antagonistic in quantum theory, and it is argued that a better
feedback control scheme should reach a trade-off between them.
Recently the proposed scheme of weak measurement-based feedback
control has focused on achieving a balance for battling decoherence
by setting proper measurement strength \cite{24,25,26}.

Recently there is a new trend in exploiting unknown quantum states to implement an advanced quantum communication and distributed quantum computation.
For instance, quantum communication of an unknown state from a photon to a photon and a blind quantum computation have been demonstrated experimentally  \cite{27,28}.
So how to protect an unknown quantum state against decoherence is of great significance.
Most of previous works on quantum state protection using quantum measurement and
feedback control have focused on protocols for known states \cite{29},
and few of them has addressed the issue of protecting unknown
quantum states. It was suggested in Ref.~\cite{30} that there might not be any effective feedback control scheme to protect an unknown state and it was argued that because we do not know any information about the state of the system before the measurement, so after measurement we just obtain parts of information about the state,
and then do not know how to do exactly. However, in Ref.~\cite{31} the authors have presented a scheme for protecting an unknown state of a qubit by weak measurement and measurement reversal.

It is well known that all existing quantum cryptosystems use
nonorthogonal states as the carriers of information. Nonorthogonal
states cannot be cloned (duplicated) by an eavesdropper. As a
result, any eavesdropping attempt must introduce errors in the
transmission, and it can be detected by the legal users of the
communication channel. For example, two nonorthogonal states are
used to encode information in the well-known B92 quantum key
distribution protocol \cite{32}. But the serious decoherence of
commercial optical fibers which are widely used in quantum
communication greatly limits the transmission distance. There is no
doubt that protecting nonorthogonal states has great practical
significance. Recently, feedback control schemes with weak
measurement have been suggested for protecting nonorthogonal states
\cite{25,26,30,33}. In Ref.~\cite{25}, a qubit prepared in one
of two nonorthogonal states and subsequently subjected to dephasing
noise was discussed, and it was found that a quantum control scheme
based on weak measurement with an appropriate measurement strength
could realize the optimal recovery from the noise. However, these
weak measurement-based feedback control schemes strongly depend on
the nonorthogonal states to be protected, and the protecting effect
is poor.

Now we reconsider feedback control with weak measurement from a new
point of view and propose a novel scheme of feed-forward and its reversal against decoherence.
It is noted that in previous schemes the intention for measurement
is to acquire information of the protected states of system, and the
feedback controls are applied to recover the state to the initial
state based on the measurement result. In this paper, we exploit
another effect of the measurement and feedback, i.e., they can transform any initial state into any given target state \cite{30}. The procedure of our scheme is like this: Before the
noise channel a pre-weak measurement is made and according to different measurement results an operation is applied in order to transform the protected state to some state (such as the states
almost immune to the noise channel) and after the noise channel one can restore the
state to the initial state with the
reversed operations and weak measurement. Such a procedure
is an example of feedback, which in this case is
referred to as feed-forward, because the result of the operation
implemented based on the result of the pre-weak measurement before the noise channel affects the dynamical behaviors of the system in the noise channel.
In detail, our aim for pre-weak measurement and feed-forward is to
intentionally drive the qubit close to its ground state before the
noise channel, which can be thought of as a partial wave function
collapse, and to restore the state to the initial state after the
noise channel. Our scheme is universal for all initial states, i.e., all the initial states can be well protected by our scheme, but the protecting effect, i.e., the fidelity, is different and state-dependent. This means that  we do not pay attention to the protected state in the whole process, and only concerns the noise channel in contrast to other schemes. Our scheme is
consisted of weak and complete measurement before noise channel
(pre-weak measurement) and reversed weak and incomplete (partial)
measurements which select measurement result after noise channel
(post-weak measurement) respectively. Meanwhile, according to
different pre-weak measurement results, we add different feed-forward operation
and its reversal into our scheme.

By using our scheme, we show that an unknown state, two nonorthogonal states and entangled state of two qubits can all be well protected probabilistically. For an arbitrary unknown state $|\psi_0\rangle=\alpha|0\rangle+\beta|1\rangle$, the protecting effect of our scheme is better than the scheme with weak measurement and measurement reversal in Ref.~\cite{31}.
For instance, for all the quantum states with $|\beta|\geqslant|\alpha|$ and heavy amplitude damping noise,
which are easier to cause the decoherence, our scheme can possess obviously superiority over that of Ref.~\cite{31}. Particularly,
for protecting two nonorthogonal states, we can find some nonorthogonal states which can be perfectly protected in an almost deterministic way, while in Ref.~\cite{25}, the scheme with a weak measurement followed by a feedback control depends much on the initial nonorthogonal states, and the effect of protection is poor. At last, compared with Ref.~\cite{34} in which the weak measurement and measurement reversal have been used for protecting entangled state of two qubits, our scheme can achieve higher success probability for preserving the same amount of entanglement and particularly our scheme can completely circumvent entanglement sudden death (ESD) from decoherence.

The paper is structured as follows. In the next section, we introduce our scheme of weak-measurement-based feed-forward control.
In Sec. III, we show how to use our scheme to protect an unknown state. In Sec. IV, we show how to use our scheme to protect two nonorthogonal states. In Sec. V, we show how to use our scheme to protect an entangled state of two qubits. In Sec. VI, we discuss the experimental feasibility of our scheme. At last, conclusion and discussion are given in Sec. VII.

%%%%%%%%%%%%%%%%%%%%%%%%%%%%%%%%%%%%%%%%%%%%%%%%%%%%%%%%%%%%%%%%%%%%%%%%%%%%%%%%%%%%%%%%%%%%%%%%%%%%%%%%%%%%%%%%%%%%%%%%%%%%%%%%%%%%%%%%%%%%%%%%%%%%
%%%%%%%%%%%%%%%%%%%%%%%%%%%%%%%%%%%%%%%%%%%%%%%%%%%%%%%%%%%%%%%%%%%%%%%%%%%%%%%%%%%%%%%%%%%%%%%%%%%%%%%%%%%%%%%%%%%%%%%%%%%%%%%%%%%%%%%%%%%%%%%%%%%%
\section{\label{sec:level11}weak-measurement-based feed-forward control SCHEME\protect\\}

Now we introduce our weak-measurement-based feed-forward control scheme. In this paper, we only consider amplitude damping noise (AD) channel,
which can be described as:
\begin{equation}
E_1=
\begin{pmatrix}
1 & 0 \\
0 & \sqrt{1-r}
\end{pmatrix},
\\
E_2=
\begin{pmatrix}
0 & \sqrt{r} \\
0 & 0
\end{pmatrix}
\label{eq:one},
\end{equation}
where $r$ is the magnitude of decoherence. Consider a qubit initially prepared in any state $\rho$, the decoherence effect of the channel can be represented as
\begin{equation}
\rho_\varepsilon=\sum_{i=1}^2 E_i\rho E_i^+
\label{eq:one}.
\end{equation}
Our scheme of protecting a quantum state is consisted of one complete pre-weak measurement, two incomplete post-weak measurements and  two feed-forward operations and their reversals, which is
illustrated in Fig. 1.
Before the noise channel the pre-weak measurement is chosen as $\Pi_1=M_1^+M_1$ and $\Pi_2=M_2^+M_2$,
and $M_1$ and $M_2$ are represented respectively as
\begin{equation}
M_1=
\begin{pmatrix}
\sqrt{p} & 0 \\
0 & \sqrt{1-p}
\end{pmatrix},
\\
M_2=
\begin{pmatrix}
\sqrt{1-p} & 0 \\
0 & \sqrt{p}
\end{pmatrix}
\label{eq:one},
\end{equation}
where the pre-weak measurement is a complete measurement, i.e., $I=\sum_{i=1}^2 M_i^+ M_i$ and $p$ is the pre-weak measurement strength.
\par
The scheme for protecting quantum state is demonstrated as follows.
For any initial state, first we make the pre-weak measurement. If
result 1 (corresponding to $\Pi_1=M_1^+M_1$) is acquired, we choose
a feed-forward operation before the noise channel, do-nothing $F_1$, i.e.,
\begin{equation}
F_1=
\begin{pmatrix}
1 & 0 \\
0 & 1
\end{pmatrix}
\label{eq:one},
\end{equation}
and then let the qubit enter the decoherence channel. Consequently,
after the noise channel a reversed operation we choose again is do-nothing $F_1$. At last, we measure
the qubit using partial weak measurement namely post-weak
measurement $\Lambda=N_1^+N_1$, and $N_1$  is represented as
\begin{equation}
N_1=
\begin{pmatrix}
\sqrt{1-p_u} & 0 \\
0 & 1
\end{pmatrix}
\label{eq:one},
\end{equation}
where the post-weak measurement is weak and incomplete and $p_u$ is the post-weak measurement strength.
\par
If result 2 (corresponding to $\Pi_2=M_2^+M_2$) of the pre-weak measurement is acquired,
we choose feed-forward operation $F_2$ as $\sigma_x$, i.e.,
\begin{equation}
F_2=
\begin{pmatrix}
0 & 1 \\
1 & 0
\end{pmatrix}
\label{eq:one},
\end{equation}
and then let the qubit enter the decoherence channel.
Consequently, after the noise channel the reversed operation we choose again is operation $F_2$, and it is noted that $F_2^2=I$.
At last, we measure the qubit using post-weak measurement $\Xi=W_1^+W_1$,
and $W_1$ is defined as
\begin{equation}
W_1=
\begin{pmatrix}
1 & 0 \\
0 & \sqrt{1-p_v}
\end{pmatrix}
\label{eq:one},
\end{equation}
where the post-weak measurement is also weak and incomplete and $p_v$ is the post-weak measurement strength. It is noted that the pre-weak measurement rather than projected measurement is necessary for our scheme because in this case the information about the initial quantum state can be preserved. And the post-weak measurement $N_1$ ($W_1$) is also very important, which is approximately the reversal of the corresponding pre-weak measurement $M_1$ ($M_2$), i.e., $M_1N_1\sim I$ and $M_2W_1\sim I$, and only in this way the information about the initial state can be retrieved. And after all these operations the original quantum state can be recovered. For clarity, we emphasize that in this paper we will frequently use these defined operators of measurements and feed-forward operations in the following sections.

%\begin{widetext}
\begin{figure}
\centering
\vspace{-2.5cm}
\includegraphics[width=0.50\textwidth]{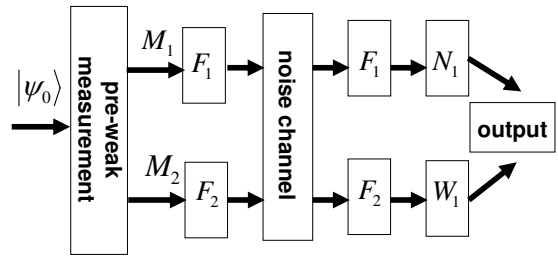}
\vspace{-5.5cm}
% Here is how to import EPS art
\caption{\label{fig:epsart} A schematic diagram showing the procedure of our feed-forward control scheme for protecting an initial state. A qubit initially prepared in an arbitrary state $|\psi_0\rangle$, and before the noise channel a complete pre-weak measurement $\{ M_1, M_2\}$ is made. If result 1 (corresponding to $M_1$) is acquired, a feed-forward operation do-nothing $F_1$ is applied. And the qubit is allowed to enter the noise channel and after the noise channel a reversed operation also do-nothing $F_1$ is applied, and then a post-weak incomplete measurement $N_1$ is made. If result 2 (corresponding to $M_2$) is acquired, a feed-forward operation $F_2$ is applied. And the qubit is allowed to enter the noise channel and after the noise channel a reversed operation also $F_2$ is applied, and then a post-weak incomplete measurement $W_1$ is made. At last the output state, which is close to the initial state, is obtained.}
\end{figure}
%\vspace{0.0cm}
%\end{widetext}

%%%%%%%%%%%%%%%%%%%%%%%%%%%%%%%%%%%%%%%%%%%%%%%%%%%%%%%%%%%%%%%%%%%%%%%%%%%%%%%%%%%%%%%%%%%%%%%%%%%%%%%%%%%%%%%%%%%%%%%%%%%%%%%%%%%%%%%%%%%%%%%%%%%%
%%%%%%%%%%%%%%%%%%%%%%%%%%%%%%%%%%%%%%%%%%%%%%%%%%%%%%%%%%%%%%%%%%%%%%%%%%%%%%%%%%%%%%%%%%%%%%%%%%%%%%%%%%%%%%%%%%%%%%%%%%%%%%%%%%%%%%%%%%%%%%%%%%%%
\section{PROTECTING UNKNOWN QUANTUM STATES}
We first consider using our scheme to protect an arbitrary unknown quantum state.
We assume that the qubit is initially prepared in an arbitrary state that we want to protect from the noise,
\begin{equation}
\begin{split}
|\psi_0\rangle=\alpha|0\rangle+\beta|1\rangle
\label{eq:one}.
\end{split}
\end{equation}
First we measure $|\psi_0\rangle$ by using pre-weak measurement $\{M_1, M_2\}$.
Depending on different measurement results, we classify the protecting process into two cases.

Case one: If result 1 is acquired, the qubit state becomes
\begin{equation}
\begin{split}
|\psi_{0}^{m1}\rangle =\frac{M_1|\psi_0\rangle}{\sqrt{\langle\psi_0|\Pi_1|\psi_0\rangle}}
\label{eq:one}
\end{split}
\end{equation}
with the probability $g^{m1} =\langle\psi_0|\Pi_1|\psi_0\rangle=|\alpha|^2p+|\beta|^2(1-p)$.
Then we adopt the feed-forward operation of do-nothing
\begin{equation}
\begin{split}
|\psi_{0}^{f1}\rangle =& F_1|\psi_{0}^{m1}\rangle=\alpha^{f1}|0\rangle+\beta^{f1}|1\rangle\\
=&\frac{1}{\sqrt{g^{m1}}}(\alpha\sqrt{p}|0\rangle+\beta\sqrt{1-p}|1\rangle)
\label{eq:one}.
\end{split}
\end{equation}
After a period of $\tau$ in the decoherence channel, the qubit state is no longer
pure because of energy relaxation with the
rate $\Gamma$. However, it is technically easier to use the
mathematical trick of unraveling the relaxation into ``jump"
and ``no jump" scenarios and work with pure states \cite{31}.
Through the decoherence channel of the amplitude damping, the qubit trajectories can be viewed
as two parts including ``jumping" into the state $|0\rangle$ with the ``jump" probability $g^{j1}=g^{m1}|\beta^{f1}|^2(1-e^{-\Gamma\tau})$ and the evolution, i.e., ``no jumping" into the state
\begin{equation}
\begin{split}
|\psi_{0}^{e1}\rangle=\frac{1}{\sqrt{g^{e1}}}(\alpha\sqrt{p}|0\rangle+\beta\sqrt{1-p}e^{-\Gamma\frac{\tau}{2}}|1\rangle)
\label{eq:one}
\end{split}
\end{equation}
with the ``no jump" probability $g^{e1}= |\alpha|^2p+|\beta|^2(1-p)e^{-\Gamma\tau}$.
Then the reversed operation we adopt is still do-nothing. The qubit state from the ``no jumping" trajectory becomes
\begin{equation}
\begin{split}
|\psi_{0}^{u1}\rangle = F_1|\psi_{0}^{e1}\rangle=|\psi_{0}^{e1}\rangle
\label{eq:one},
\end{split}
\end{equation}
and the qubit state from the ``jumping" trajectory becomes
\begin{equation}
\begin{split}
|\psi_{0}^{ju1}\rangle = F_1|0\rangle=|0\rangle
\label{eq:one}.
\end{split}
\end{equation}
At last we measure the qubit by using the post-weak measurement $\Lambda=N_1^+N_1$ and obtain the measured state from the ``no jumping" trajectory
\begin{equation}
\begin{split}
|\psi_{0}^{n1}\rangle =&\frac{N_1|\psi^{u1}_0\rangle}{\sqrt{\langle\psi^{u1}_0|\Lambda|\psi^{u1}_0\rangle}}\\
=&\frac{1}{\sqrt{g^{n1}}}(\alpha\sqrt{p}\sqrt{1-p_u}|0\rangle+\beta\sqrt{1-p}e^{-\Gamma\frac{\tau}{2}}|1\rangle)
\label{eq:one}
\end{split}
\end{equation}
with the probability $g^{n1}=|\alpha|^2p(1-p_u)+|\beta|^2(1-p)e^{-\Gamma\tau}$.
The state $|\psi_{0}^{ju1}\rangle$ from the ``jumping" trajectory bacomes $|0\rangle$
with the probability $g^{jn1} = g^{j1} (1-p_u)= g^{m1}|\beta^{f1}|^2(1-e^{-\Gamma\tau})(1-p_u)$.
The density matrix description of the final qubit state can be written as
\begin{equation}
\begin{split}
\rho^{fin1}_0=\frac{g^{n1}|\psi_{0}^{n1}\rangle\langle\psi_{0}^{n1}|+g^{jn1}|0\rangle\langle0|}{g^{n1}+g^{jn1}}
\label{eq:one}
\end{split}
\end{equation}
with the success (selection) probability and fidelity respectively
\begin{equation}
\begin{split}
g^{fin1}=g^{n1}+g^{jn1}
\label{eq:one},
\end{split}
\end{equation}
\begin{equation}
\begin{split}
f^{fin1}=\langle\psi_{0}|\rho^{fin1}_0|\psi_{0}\rangle
\label{eq:one}.
\end{split}
\end{equation}

Case two: If we obtain result 2 in the pre-weak measurement, the qubit state becomes
\begin{equation}
\begin{split}
|\psi_{0}^{m2}\rangle =\frac{M_2|\psi_0\rangle}{\sqrt{\langle\psi_0|\Pi_2|\psi_0\rangle}}
\label{eq:one}
\end{split}
\end{equation}
with the probability $g^{m2}=\langle\psi_0|\Pi_2|\psi_0\rangle =|\alpha|^2(1-p)+|\beta|^2p$, and we adopt the feed-forward operation of $F_2$
\begin{equation}
\begin{split}
|\psi_{0}^{f2}\rangle =& F_2|\psi_{0}^{m2}\rangle=\alpha^{f2}|0\rangle+\beta^{f2}|1\rangle\\
=&\frac{1}{\sqrt{g^{m2}}}(\beta\sqrt{p}|0\rangle+\alpha\sqrt{1-p}|1\rangle)
\label{eq:one}.
\end{split}
\end{equation}
Then like case one, we let the qubit go through the decoherence channel and we also view the qubit trajectories as two parts ``jumping" into the state $|0\rangle$ with the ``jump" probability $g^{j2}=g^{m2}|\beta^{f2}|^2(1-e^{-\Gamma\tau})$ and the evolution, i.e., ``no jumping" into the state
\begin{equation}
\begin{split}
|\psi_{0}^{e2}\rangle=&\frac{1}{\sqrt{g^{e2}}}(\beta\sqrt{p}|0\rangle+\alpha\sqrt{1-p}e^{-\Gamma\frac{\tau}{2}}|1\rangle)
\label{eq:one}
\end{split}
\end{equation}
with the probability $g^{e2}= |\beta|^2p+|\alpha|^2(1-p)e^{-\Gamma\tau}$.
After the noise channel the reversed operation we apply is still $F_2$, and the qubit state of the ``no jumping" trajectory becomes
\begin{equation}
\begin{split}
|\psi_{0}^{u2}\rangle =& F_2|\psi_{0}^{e2}\rangle\\
=&\frac{1}{\sqrt{g^{e2}}}(\alpha\sqrt{1-p}e^{-\Gamma\frac{\tau}{2}}|0\rangle+\beta\sqrt{p}|1\rangle)
\label{eq:one},
\end{split}
\end{equation}
and the qubit state of the ``jumping" trajectory becomes
\begin{equation}
\begin{split}
|\psi_{0}^{ju2}\rangle = F_2|0\rangle=|1\rangle
\label{eq:one}.
\end{split}
\end{equation}
At last we measure the qubit by using the post-weak measurement $\Xi=W_1^+W_1$ and obtain the following state from the ``no jumping" trajectory
\begin{equation}
\begin{split}
|\psi_{0}^{w1}\rangle =&\frac{W_1|\psi^{u2}_0\rangle}{\sqrt{\langle\psi^{u2}_0|\Xi|\psi^{u2}_0\rangle}}\\
=&\frac{1}{\sqrt{g^{w1}}}(\alpha\sqrt{1-p}e^{-\Gamma\frac{\tau}{2}}|0\rangle+\beta\sqrt{p}\sqrt{1-p_v}|1\rangle)
\label{eq:one}
\end{split}
\end{equation}
with the probability $g^{w1}=|\alpha|^2(1-p)e^{-\Gamma\tau}+|\beta|^2p(1-p_v)$.
The state $|\psi_{0}^{ju2}\rangle$ from the ``jumping" trajectory becomes $|1\rangle$
with the probability $g^{jw1} = g^{j2}(1-p_v)= g^{m2}|\beta^{f2}|^2(1-e^{-\Gamma\tau})(1-p_v)$.
The density matrix description of the final qubit state can be written as
\begin{equation}
\begin{split}
\rho^{fin2}_0=\frac{g^{w1}|\psi_{0}^{w1}\rangle\langle\psi_{0}^{w1}|+g^{jw1}|1\rangle\langle1|}{g^{w1}+g^{jw1}}
\label{eq:one}
\end{split}
\end{equation}
with the success (selection) probability and fidelity respectively
\begin{equation}
\begin{split}
g^{fin2}=g^{w1}+g^{jw1}
\label{eq:one},
\end{split}
\end{equation}
\begin{equation}
\begin{split}
f^{fin2}=\langle\psi_{0}|\rho^{fin2}_0|\psi_{0}\rangle
\label{eq:one}.
\end{split}
\end{equation}
\par
After completing the whole measurement and feed-forward process,  the final total success probability and the final total fidelity can be expressed respectively as
\begin{equation}
\begin{split}
g^{fin}=g^{fin1}+g^{fin2}
\label{eq:one},
\end{split}
\end{equation}

\begin{equation}
\begin{split}
f^{fin}=\frac{g^{fin1} f^{fin1}+g^{fin2} f^{fin2}}{g^{fin}}
\label{eq:one}.
\end{split}
\end{equation}
It is noted that if we let $1-e^{-\Gamma\tau}=r$, the decoherence channel described in this section is the same as the channel mentioned in section II.

\subsection{\label{sec:level2} Exact protection}
It is noted that in case one if we take the post-weak measurement strength as
\begin{equation}
\begin{split}
p_u=1-\frac{(1-p)e^{-\Gamma\tau}}{p}=1-\frac{(1-p)(1-r)}{p}
\label{eq:one},
\end{split}
\end{equation}
the measured qubit state $|\psi_{0}^{n1}\rangle$ in Eq.~(14) along the ``no jumping" trajectory will be the initial state, i.e.,  $|\psi_{0}^{n1}\rangle=|\psi_{0}\rangle$, which would be an exact restoration of the initial state, and we define it as an exact protection.
And analogously, in case two if we take the weak measurement strength
\begin{equation}
\begin{split}
p_v=1-\frac{(1-p)e^{-\Gamma\tau}}{p}=1-\frac{(1-p)(1-r)}{p}
\label{eq:one},
\end{split}
\end{equation}
the measured qubit state $|\psi_{0}^{w1}\rangle$ in Eq.~(23) along the ``no jumping" trajectory will be the initial state, i.e.,  $|\psi_{0}^{w1}\rangle=|\psi_{0}\rangle$,
which would also be an exact restoration of the initial state.
We define Eqs.~(29) and (30) as an exact protecting condition in our scheme.
It can be seen from Eqs.~(29, 30) that because  $p_u$ and $p_v\geq0$,  $p\in[\frac{1-r}{2-r}, 1]$.
Under this condition, for an arbitrary state, the final total success probability and the final total fidelity become respectively
\begin{equation}
\begin{split}
g^{fin}=&g^{fin1}+g^{fin2}\\
=&\frac{(1-p)^2r(1-r)}{p}+2(1-p)(1-r)
\label{eq:one},
\end{split}
\end{equation}
\begin{equation}
\begin{split}
f^{fin}=&\frac{g^{fin1} f^{fin1}+g^{fin2} f^{fin2}}{g^{fin}}\\
=&\frac{\frac{2(1-p)^2r(1-r)}{p}|\alpha|^2|\beta|^2}{\frac{(1-p)^2r(1-r)}{p}+2(1-p)(1-r)}\\
&+\frac{2(1-p)(1-r)}{\frac{(1-p)^2r(1-r)}{p}+2(1-p)(1-r)}
\label{eq:one}.
\end{split}
\end{equation}
We can see that in this case the final total success probability is independent of the state. However, the final total fidelity is state dependent.
An arbitrary state on the Bloch sphere can be represented in terms of $\alpha=\cos{\frac{\theta}{2}}$ and $\beta=e^{i\phi}\sin{\frac{\theta}{2}}$. If we take an average over the whole Bloch sphere, we obtain the average fidelity $f_{ave}$
\begin{equation}
\begin{split}
f_{ave}=&\frac{1}{4\pi}\int_0^{2\pi}\int_0^{\pi}f^{fin}\sin\theta d\theta d\phi\\
=&\frac{\frac{(1-p)^2r(1-r)}{3p}+2(1-p)(1-r)}{\frac{(1-p)^2r(1-r)}{p}+2(1-p)(1-r)}
\label{eq:one}.
\end{split}
\end{equation}

\begin{widetext}

\begin{figure}[t]
\centering
\vspace{-2.2cm}
\includegraphics[scale=0.77]{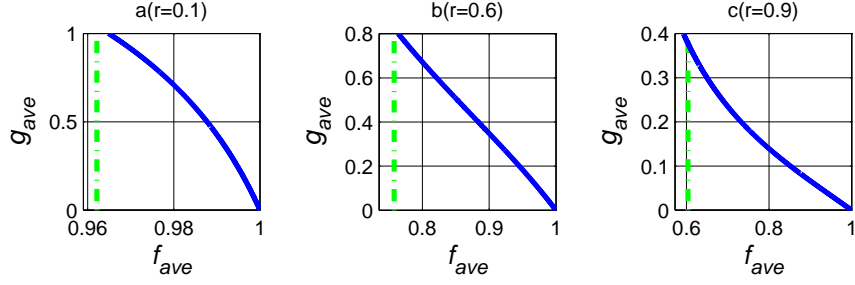}%{unn812.pdf}%{pic43.pdf}%{ppm813.pdf}%{pic43.pdf}%{pictu82.pdf}
\vspace{-2.4cm}
% Here is how to import EPS art
\caption{\label{fig:epsart} (Color online). Under the exact protecting condition, $f_{ave}$-$g_{ave}$ phase diagrams for a: r=0.1, b: r=0.6 and c: r=0.9.
The green dotted lines represent the fidelity without any measurement and feed-forward.}
\vspace{0.6cm}
\end{figure}
\vspace{0.0cm}
\end{widetext}
It is noted that the average success probability is still $g^{fin}$ due to its independence of the state, but for clarity, we still define a symbol $g_{ave}$ ($=g^{fin}$) as the average success probability. From numerical calculation we find that the fidelity will increase with the measurement strength $p$, and
when we choose $p$ close to 1, the fidelity will approach 1, that means we can make the final qubit state arbitrarily
close to the initial state under the exact protection. On the contrary, the success probability will decrease with the measurement strength $p$,
when we choose $p$ close to 1, the success probability almost drops to 0.

Under the exact protecting condition, the relation between the average fidelity $f_{ave}$ and the average success probability $g_{ave}$ is shown in Fig.~2. As an example, we only show the protecting effect of an unknown state for $r=0.1, 0.6$ and $0.9$.
It can be seen that our scheme greatly improves the fidelity than that without any measurement and feed-forward. Especially, our scheme can protect an arbitrary unknown state even for heavy amplitude damping noise (e.g., r=0.9).

\subsection{\label{sec:level2} Optimal protection}
In part A, we discuss the exact protection of an arbitrary unknown state,
which demands $p_u$ and $p_v$ having specific relationship with $p$.
However, this kind of protection is not an optimal protection.
Actually all measurement strength $p, p_u$ and $p_v$ can be varied independently from 0 to 1 so that by varying them we can obtain the highest success probability (fidelity) for fixed fidelity (success probability) for an arbitrary unknown state. We call this kind of protection as optimal protection and the corresponding groups of measurement strength $p, p_u$ and $p_v$ as optimal protecting condition.
Now we show the optimal protection of an unknown state by using our scheme.
According to Eqs.~(27) and (28), for an arbitrary state,
the final total success probability and the final total fidelity can be expressed respectively as
\begin{widetext}

\begin{equation}
\begin{split}
G^{fin}=&G^{fin1}+G^{fin2}\\
=&p(1-p_u)|\alpha|^2+(1-p)(1-p_u)r|\beta|^2+p(1-p_v)|\beta|^2\\
&+(1-p)(1-p_v)r|\alpha|^2+(1-p)(1-r)
\label{eq:one},
\end{split}
\end{equation}

\begin{equation}
\begin{split}
F^{fin}=&\frac{G^{fin1} F^{fin1}+G^{fin2} F^{fin2}}{G^{fin}}\\
=&\frac{p(1-p_u)|\alpha|^4+(1-p)(1-r)|\beta|^4+
(1-p)(1-p_u)r|\alpha|^2|\beta|^2+2\sqrt{p}\sqrt{1-p}\sqrt{1-p_u}\sqrt{1-r}|\alpha|^2|\beta|^2}
{p(1-p_u)|\alpha|^2+(1-p)(1-p_u)r|\beta|^2+p(1-p_v)|\beta|^2+(1-p)(1-p_v)r|\alpha|^2+(1-p)(1-r)}+\\
&\frac{(1-p)(1-r)|\alpha|^4+p(1-p_v)|\beta|^4+
(1-p)(1-p_v)r|\alpha|^2|\beta|^2+2\sqrt{p}\sqrt{1-p}\sqrt{1-p_v}\sqrt{1-r}|\alpha|^2|\beta|^2}
{p(1-p_u)|\alpha|^2+(1-p)(1-p_u)r|\beta|^2+p(1-p_v)|\beta|^2+(1-p)(1-p_v)r|\alpha|^2+(1-p)(1-r)}
\label{eq:one}.
\end{split}
\end{equation}
\end{widetext}
An arbitrary state on the Bloch sphere can be represented in terms of $\alpha=\cos{\frac{\theta}{2}}$ and $\beta=e^{i\phi}\sin{\frac{\theta}{2}}$.
For an arbitrary state, it is clear that each group of measurement strength ($p, p_u, p_v$) uniquely determines its corresponding fidelity and success probability denoted by one point ($F^{fin}$, $G^{fin}$).
When the measurement strengths $p, p_u, p_v$ are independently taken over all the real numbers from 0 to 1, we obtain $F^{fin}$ (the fidelity)-$G^{fin}$ (success probability) phase diagram. We first show the protecting effect of several states with $\theta=\frac{3\pi}{8}, \frac{\pi}{2}$, $\frac{3\pi}{4}$ and $\phi=0$ for $r=0.6$ in Fig.~3.
\begin{widetext}

\begin{figure}
\vspace{0.3cm}
\includegraphics[scale=0.55]{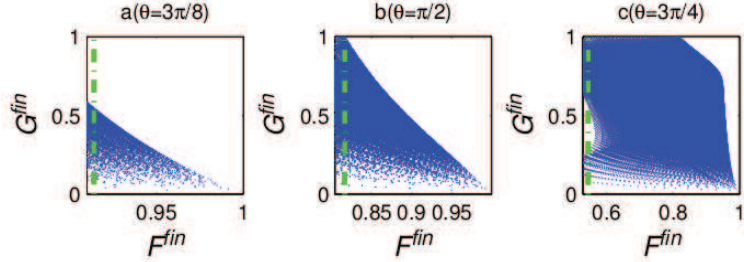}%{unn954.pdf}%{unn951.pdf}%{ppm95.pdf}%{pp932.pdf}%{p442.pdf}%{pic93.pdf}% Here is how to import EPS art
\vspace{-0.1cm}
\centering
\caption{\label{fig:epsart}(Color online). $F^{fin}$-$G^{fin}$ phase diagrams for protecting some given states a: $\theta=\frac{3\pi}{8}$, b: $\theta=\frac{\pi}{2}$ and c: $\theta=\frac{3\pi}{4}$ . We set $r=0.6$ and $\phi=0$ for three diagrams. The optimal protection is given by the boundary lines of the blue region.
The green dotted lines represent the fidelity without any measurement and feed-forward.}
\vspace{0.2cm}
\end{figure}
\end{widetext}
It is noted that for given fidelity (succuss probability), the point ($F^{fin}$, $G^{fin}$), at which the succuss probability (fidelity) is maximized, is distributed on the boundary of the phase diagram. All the points distributed on the boundary line of the phase diagram correspond to the optimal protection and we denote these special points by the optimal fidelity $F^{fin}_{opt}$ and the optimal success probability $G^{fin}_{opt}$.
These groups of measurement strength ($p, p_u, p_v$) corresponding to the optimal protection points ($F^{fin}_{opt}$, $G^{fin}_{opt}$) on the boundary line are  optimal groups of measurement strength (i.e., optimal protecting condition).
The boundary line of the phase diagram indicates the relationship between the fidelity and the succuss probability under optimal protecting condition.
From Fig.~3 we can see that under the optimal protecting condition, the larger $|\beta|/|\alpha|$ becomes, the more superiority our scheme has, which is very important in quantum communication due to these states being much easier to be influenced by the noise channel.
In Ref.~\cite{29} the optimal quantum feedback control based on weak measurement is available for all typical types of noise sources,
but the approach is valid only for some suitable quantum states and the fidelity has only an improvement less than three percent compared with that without any measurement and feedback. Compared with Ref.~\cite{29}, our scheme can probabilistically protect an arbitrary state with higher fidelity.

Now we consider optimal protection of an unknown state. By taking average over the whole Bloch sphere, the average success probability $G^{ave}$ and the average fidelity $F^{ave}$ can be expressed respectively as
\begin{equation}
\begin{split}
G^{ave}=\frac{1}{4\pi}\int_0^{2\pi}\int_0^{\pi}G^{fin}\sin\theta d\theta d\phi
\label{eq:one}.
\end{split}
\end{equation}
\begin{equation}
\begin{split}
F^{ave}=\frac{1}{4\pi}\int_0^{2\pi}\int_0^{\pi}F^{fin}\sin\theta d\theta d\phi
\label{eq:one}.
\end{split}
\end{equation}

%\begin{widetext}

\begin{figure}
\centering
\vspace{0.3cm}
\includegraphics[scale=0.45]{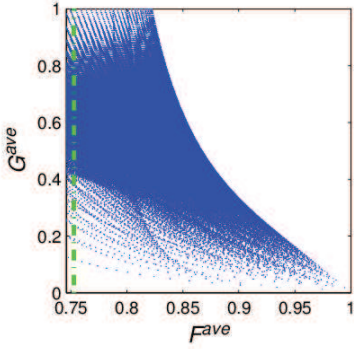}%{mypuvv4.pdf}%{mypuv102.pdf}%{mypuvm.pdf}%{mypuvall002.pdf}%{mypuvall01.pdf}
\vspace{-0.1cm}
% Here is how to import EPS art
\caption{\label{fig:epsart} (Color online). $F^{ave}$-$G^{ave}$ phase  diagram.
The green dotted line represents the fidelity without any measurement and feed-forward.}
\end{figure}
%\end{widetext}
For $r=0.6$, we plot $F^{ave}$- $G^{ave}$ phase diagram in Fig.~4.
Similarly we can obtain the optimal protection and corresponding condition in this case. The average optimal protection points ($F^{ave}_{opt}$, $G^{ave}_{opt}$) with the average fidelity $F^{ave}_{opt}$ and the average succuss probability $G^{ave}_{opt}$ under the optimal protecting condition lie on the boundary of the phase diagram.
It can be seen from Fig.~4 that the average fidelity can be much larger than that without measurement and feed-forward (green dotted line), even close to 1. Of course, it is at the cost of the decrease of the success probability.
However, compared with the scheme of Ref.~\cite{31} based on weak measurement and measurement reversal without feed-forward control, our scheme can reach higher success probability for fixed fidelity. In addition, we find that our scheme is obviously superior over that of Ref.~\cite{31} at any degree of amplitude damping, and the heavier the amplitude damping noise, the better protecting effect of our scheme than that of their scheme. As an example, under the optimal protecting condition, we numerically compare our scheme (boundary lines of the blue region) with that in Ref.~\cite{31} (boundary lines of the red region) for $r=0.1, 0.6$ and $0.9$ in Fig.~5.
We can see that under the optimal protection, our scheme is always better than that in Ref.~\cite{31}.
In particular, when the amplitude damping is large such as $r=0.9$, our scheme has an obvious advantage.
The feature of our scheme has practical significance in current communication, due to larger decoherence in commercial optical fibers with long transmission distance.
\begin{widetext}

\begin{figure}
\centering
\vspace{0.3cm}
\includegraphics[scale=0.55]{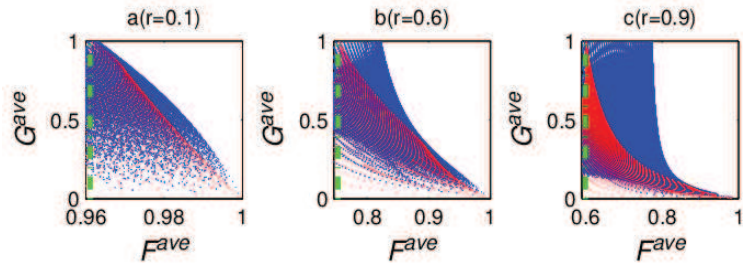}%{unave1023.pdf}%{ppm1021.pdf}%{pictu.pdf}% Here is how to import EPS art
\vspace{-0.1cm}
\caption{\label{fig:epsart}(Color online). $F^{ave}$-$G^{ave}$ phase diagrams for different amplitude damping a: r=0.1, b: r=0.6 and c: r=0.9.
The blue region is corresponding to the results of our scheme, and the red region is corresponding to that of Ref.~\cite{31}.
The green dotted lines represent the fidelity without any measurement and feed-forward.}
\end{figure}
\end{widetext}

%%%%%%%%%%%%%%%%%%%%%%%%%%%%%%%%%%%%%%%%%%%%%%%%%%%%%%%%%%%%%%%%%%%%%%%%%%%%%%%%%%%%%%%%%%%%%%%%%%%%%%%%%%%%%%%%%%%%%%%%%%%%%%%%%%%%%%%%%%%%%%%%%%%%
%%%%%%%%%%%%%%%%%%%%%%%%%%%%%%%%%%%%%%%%%%%%%%%%%%%%%%%%%%%%%%%%%%%%%%%%%%%%%%%%%%%%%%%%%%%%%%%%%%%%%%%%%%%%%%%%%%%%%%%%%%%%%%%%%%%%%%%%%%%%%%%%%%%%
\section{PROTECTING TWO NONORTHOGONAL STATES}
\vspace{-0.2cm}
In previous section, we show how to use our scheme to protect an unknown quantum state, and in this section we will show that it is more effective to protect two known nonorthogonal states. It is noted that the procedure of protecting each one of two nonorthogonal states is the same as that of protecting an arbitrary state  mentioned in section III.
Consider two nonorthogonal states that we want to protect from noise,
\begin{equation}
\begin{split}
|\psi_\pm\rangle=&\cos{\frac{\theta}{2}}|+\rangle\pm e^{i\phi}\sin{\frac{\theta}{2}}|-\rangle\\
=&\frac{1}{\sqrt{2}}(\cos{\frac{\theta}{2}}\pm e^{i\phi}\sin{\frac{\theta}{2}})|0\rangle+\\
&\frac{1}{\sqrt{2}}(\cos{\frac{\theta}{2}}\mp e^{i\phi}\sin{\frac{\theta}{2}})|1\rangle\\
=&\alpha_\pm|0\rangle+\beta_\pm|1\rangle
\label{eq:one},
\end{split}
\end{equation}
where $|\pm\rangle= \frac{1}{\sqrt{2}}(|0\rangle\pm|1\rangle)$,
$\alpha_\pm=\frac{1}{\sqrt{2}}(\cos{\frac{\theta}{2}}\pm
e^{i\phi}\sin{\frac{\theta}{2}})$ and
$\beta_\pm=\frac{1}{\sqrt{2}}(\cos{\frac{\theta}{2}}\mp
e^{i\phi}\sin{\frac{\theta}{2}})$. It is noted that the
nonorthogonal states $|\psi_\pm\rangle$ are more general than that
in Refs.~\cite{25,26}. The overlapping of the two nonorthogonal
states can be quantified by
\begin{equation}
\langle\psi_+|\psi_-\rangle= \cos\theta
\label{eq:one},
\end{equation}
which is independent of $\phi$. The range of $\theta$ and $\phi$ is respectively from $0$  to $\frac{\pi}{2}$ and from $0$ to $2\pi$.\\

\subsection{\label{sec:level2}Exact protection}
Under the exact protecting condition $p_u=1-\frac{(1-p)(1-r)}{p}$ and $p_v=1-\frac{(1-p)(1-r)}{p}$,
the final total success probability and the final total fidelity of each one of two nonorthogonal states can be obtained from Eqs.~(31) and (32)
\begin{widetext}

\begin{equation}
\begin{split}
g^{fin}_\pm=\frac{(1-p)^2r(1-r)}{p}+2(1-p)(1-r)
\label{eq:one},
\end{split}
\end{equation}

\begin{equation}
\begin{split}
f^{fin}_\pm=\frac{\frac{2(1-p)^2r(1-r)}{p}|\alpha_\pm|^2|\beta_\pm|^2}{\frac{(1-p)^2r(1-r)}{p}+2(1-p)(1-r)}
+\frac{2(1-p)(1-r)}{\frac{(1-p)^2r(1-r)}{p}+2(1-p)(1-r)}
\label{eq:one}.
\end{split}
\end{equation}
Substituting  $\alpha_\pm=\frac{1}{\sqrt{2}}(\cos{\frac{\theta}{2}}\pm e^{i\phi}\sin{\frac{\theta}{2}})$ and $\beta_\pm=\frac{1}{\sqrt{2}}(\cos{\frac{\theta}{2}}\mp e^{i\phi}\sin{\frac{\theta}{2}})$ into Eq.~(41), we obtain
\begin{equation}
\begin{split}
f^{fin}_\pm=\frac{\frac{(1-p)^2r(1-r)}{2p}|\cos\frac{\theta}{2}\pm e^{i\phi}\sin\frac{\theta}{2}|^2|\cos\frac{\theta}{2}\mp e^{i\phi}\sin\frac{\theta}{2}|^2}{\frac{(1-p)^2r(1-r)}{p}+2(1-p)(1-r)}
+\frac{2(1-p)(1-r)}{\frac{(1-p)^2r(1-r)}{p}+2(1-p)(1-r)}
\label{eq:one}.
\end{split}
\end{equation}

\begin{figure}[t]
\centering
\vspace{-1.5cm}
\includegraphics[scale=0.75]{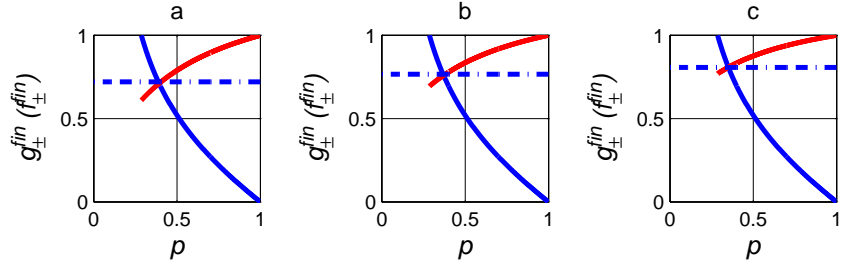}%{nonn021.pdf}%{vnon021.pdf}%{pic0021.pdf}
\vspace{-2.3cm}
\caption{\label{fig:epsart} (Color online). Under the exact protecting condition, the effect of protecting two nonorthogonal states for
a: $\theta=\frac{3\pi}{8},\phi=\frac{\pi}{16}$;   b: $\theta=\frac{\pi}{4},\phi=\frac{9\pi}{8}$;  c: $\theta=\frac{\pi}{8},\phi=\frac{7\pi}{4}$ and $r=0.6$.
The blue (red) solid lines represent the relationship between the success probability (the fidelity) and measurement strength $p$. The blue dotted lines represent the fidelity without any measurement and feed-forward.}
\end{figure}
\end{widetext}

It can be seen from Eqs.~(40) and (42) that the success probability and the fidelity are the same for both $|\psi_\pm\rangle$. Particularly, the success probability does not depend on the nonorthogonal states.
And for each one of the two nonorthogonal states,  as $p$ increases,
the fidelity increases, and on the contrary the success probability decreases.
As an example we give the protecting effect for different pairs of nonorthogonal states ($|\psi_+\rangle$,  $|\psi_-\rangle$) with $(\theta, \phi)=(\frac{3\pi}{8}, \frac{\pi}{16}), (\frac{\pi}{4}, \frac{9\pi}{8}), (\frac{\pi}{8}, \frac{7\pi}{4})$ for $r=0.6$ in Fig.~6.
The trade-off relation between the success probability and the fidelity is clearly shown in Fig.~6.
Although, the protecting effect for different nonorthogonal states are slightly different, it is obvious that the application of our scheme increases the fidelity comparing with that without any measurement and feed-forward scheme. It should be emphasized that our scheme can effectively protect any pair of two nonorthogonal states, i.e., for any $\theta$ and $\phi$ on the Bloch sphere under the exact protecting condition.\\

\subsection{\label{sec:level2} Optimal protection}
In part A, we discuss the exact protection of two nonorthogonal states,
and in this part, we show optimal protection of two nonorthogonal states.
It is noted that the average optimal protection of two nonorthogonal states is very similar to the average optimal protection of an unknown state.
According to Eqs.~(34) and (35), we can obtain the succuss probability and the fidelity for each one of two nonorthogonal states, which can be expressed respectively as

\begin{widetext}

\begin{equation}
\begin{split}
G_\pm^{fin}=&G_\pm^{fin1}+G_\pm^{fin2}\\
=&p(1-p_u)|\alpha_\pm|^2
+(1-p)(1-p_u)r|\beta_\pm|^2+p(1-p_v)|\beta_\pm|^2
+(1-p)(1-p_v)r|\alpha_\pm|^2+(1-p)(1-r)
\label{eq:one},
\end{split}
\end{equation}

\begin{equation}
\begin{split}
F^{fin}_\pm=&\frac{G^{fin1}_\pm F^{fin1}_\pm+G^{fin2}_\pm F^{fin2}_\pm}{G^{fin}_\pm}\\
=&\frac{p(1-p_u)|\alpha_\pm|^4+(1-p)(1-r)|\beta_\pm|^4+
(1-p)(1-p_u)r|\alpha_\pm|^2|\beta_\pm|^2+2\sqrt{p}\sqrt{1-p}\sqrt{1-p_u}\sqrt{1-r}|\alpha_\pm|^2|\beta_\pm|^2}
{p(1-p_u)|\alpha_\pm|^2+(1-p)(1-p_u)r|\beta_\pm|^2+p(1-p_v)|\beta_\pm|^2+(1-p)(1-p_v)r|\alpha_\pm|^2+(1-p)(1-r)}+\\
&\frac{(1-p)(1-r)|\alpha_\pm|^4+p(1-p_v)|\beta_\pm|^4+
(1-p)(1-p_v)r|\alpha_\pm|^2|\beta_\pm|^2+2\sqrt{p}\sqrt{1-p}\sqrt{1-p_v}\sqrt{1-r}|\alpha_\pm|^2|\beta_\pm|^2}
{p(1-p_u)|\alpha_\pm|^2+(1-p)(1-p_u)r|\beta_\pm|^2+p(1-p_v)|\beta_\pm|^2+(1-p)(1-p_v)r|\alpha_\pm|^2+(1-p)(1-r)}
\label{eq:one}.
\end{split}
\end{equation}
\end{widetext}

Now we assume that the probability of two nonorthogonal states to be generated is equal.
The success probability and fidelity for two nonorthogonal states can be written respectively as
\begin{equation}
\begin{split}
G^{fin}_{non}=&\frac{G_+^{fin}+G_-^{fin}}{2}\\
=&p(1-p_u)+p(1-p_v)+(1-p)(1-p_u)r \\
&+(1-p)(1-p_v)r+2(1-p)(1-r)
\label{eq:one},
\end{split}
\end{equation}
\begin{equation}
\begin{split}
F^{fin}_{non}=\frac{F_+^{fin}+F_-^{fin}}{2}
\label{eq:one}.
\end{split}
\end{equation}
For two given nonorthogonal states, i.e., for given $\theta$ and $\phi$,
we can plot $F^{fin}_{non}$-$G^{fin}_{non}$ phase diagram. Similarly we can obtain the optimal protection and the corresponding condition in this case. The optimal protection points ($F^{opt}_{non}$, $G^{opt}_{non}$) with the fidelity $F^{opt}_{non}$ and the succuss probability $G^{opt}_{non}$ under the optimal protecting condition lie on the boundary of the phase diagram.
As an example, we numerically demonstrate the optimal protection of different nonorthogonal states for $r=0.6$ in Fig.~7.
\begin{widetext}

\begin{figure}[t]
\centering
\vspace{0.5cm}
\includegraphics[scale=0.55]{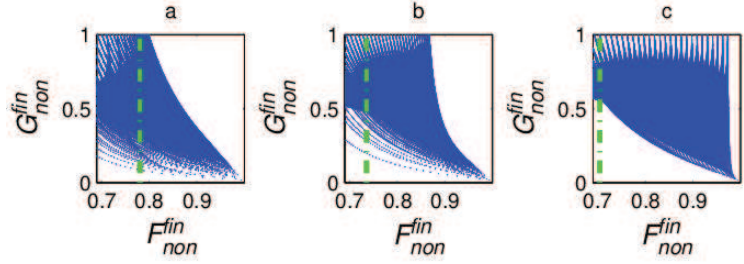}%{matchch1.pdf}%{matchch1.pdf}%{match23.pdf}%{match03.pdf}%{non2.pdf}
\vspace{0.0cm}
% Here is how to import EPS art
\centering
\caption{\label{fig:epsart} (Color online). $F^{fin}_{non}$-$G^{fin}_{non}$ phase diagrams of protecting two nonorthogonal states a: $\theta=\frac{3\pi}{16},\phi=0$;   b: $\theta=\frac{5\pi}{16},\phi=0$;  c: $\theta=\frac{7\pi}{16},\phi=0$ with $r=0.6$. The green dotted lines represent the fidelity without any measurement and feed-forward control.}
\end{figure}
\end{widetext}
We can see that our scheme can well protect two nonorthogonal states under the optimal protecting condition. Especially when $\theta=\frac{7\pi}{16}, \phi=0$,  the fidelity and the succuss probability can respectively reach $0.98$ and $0.9949$ as shown in Fig.~7(c) and the corresponding optimal group of measurement strength is $p=0.995,\  p_u=0.0052$ and $p_v=0.005$.
Compared with Ref.~\cite{25}, the protecting effect of our scheme for two nonorthogonal states is much better.
The scheme in Ref.~\cite{25} depends much on the initial nonorthogonal states and is not effective, while our scheme can effectively protect any pair of nonorthogonal states.

%%%%%%%%%%%%%%%%%%%%%%%%%%%%%%%%%%%%%%%%%%%%%%%%%%%%%%%%%%%%%%%%%%%%%%%%%%%%%%%%%%%%%%%%%%%%%%%%%%%%%%
%%%%%%%%%%%%%%%%%%%%%%%%%%%%%%%%%%%%%%%%%%%%%%%%%%%%%%%%%%%%%%%%%%%%%%%%%%%%%%%%%%%%%%%%%%%%%%%%%%%%%%%%%%%
\section{PROTECTING ENTANGLED STATE}
It is noted that our scheme can also protect multipartite quantum
state. Now we consider a two-qubit (labeled by a and b) quantum
system whose initial state is
$|\Phi\rangle=\alpha|00\rangle+\beta|11\rangle$. We assume that the
two qubits being in different environments go respectively through
two independent quantum channels with damping rates $r_a$ and $r_b$.
By using our scheme for each of the two qubits, we can protect
this entangled state of the two qubits. According to the pre-weak measurement
results for each qubit, we can classify our protecting scheme into
four cases, i.e., $M_{c1}=M_1^a\otimes M_1^b$, $M_{c2}=M_2^a\otimes
M_2^b$, $M_{c3}=M_1^a\otimes M_2^b$ and $M_{c4}=M_2^a\otimes M_1^b$,
where
\begin{equation*}
\begin{split}
M_1^i=
\begin{pmatrix}
\sqrt{p^i} & 0 \\
0 & \sqrt{1-p^i}
\end{pmatrix},
M_2^i=
\begin{pmatrix}
\sqrt{1-p^i} & 0 \\
0 & \sqrt{p^i}
\end{pmatrix}
\label{eq:one},(i=a, b).
\end{split}
\end{equation*}
It is noted that the pre-weak measurement of two qubits is complete,
i.e., $I=\sum_{i=1}^4M_{ci}M_{ci}^+$. We define the feed-forward
operations as
\begin{equation}
\begin{split}
F_1^a = F_1^b=
\begin{pmatrix}
1 & 0 \\
0 & 1
\end{pmatrix},
F_2^a = F_2^b=
\begin{pmatrix}
0 & 1 \\
1 & 0
\end{pmatrix}
\label{eq:one},
\end{split}
\end{equation}
and the post-weak measurements as
\begin{equation}
\begin{split}
N_1^i=
\begin{pmatrix}
\sqrt{1-p_u^i} & 0 \\
0 & 1
\end{pmatrix},
W_1^i=
\begin{pmatrix}
1 & 0 \\
0 & \sqrt{1-p_v^i}
\end{pmatrix}
\label{eq:one}, (i=a, b).
\end{split}
\end{equation}

Case one: If result 1 ($M_{c1}$) is acquired,
the feed-forward operation and its reversed operation are both chosen as $F_1^a\otimes F_1^b$
and the post-weak measurement is chosen as $N_1^a\otimes N_1^b$.

Case two: If result 2 ($M_{c2}$) is acquired,
the feed-forward operation and its reversed operation are both chosen as $F_2^a\otimes F_2^b$
and the post-weak measurement is chosen as $W_1^a\otimes W_1^b$.

Case three: If result 3 ($M_{c3}$) is acquired,
the feed-forward operation and its reversed operation are both chosen as $F_1^a\otimes F_2^b$
and the post-weak measurement is chosen as $N_1^a\otimes W_1^b$.

Case four: If result 4 ($M_{c4}$) is acquired,
the feed-forward operation and its reversed operation are both chosen as $F_2^a\otimes F_1^b$
and the post-weak measurement is chosen as $W_1^a\otimes N_1^b$.

\par In each case, the final density matrix can be written as
\begin{equation}
\begin{split}
\rho^i=\frac{1}{P_i}
\begin{pmatrix}
\rho^i_{11} & 0 &0&\rho^i_{14} \\
0 & \rho^i_{22} &0&0 \\
0& 0 &\rho^i_{33}&0 \\
\rho^i_{41} & 0 &0&\rho^i_{44} \\
\end{pmatrix}
\label{eq:one}
\end{split}
\end{equation}
with the success probability $P_i=\rho^i_{11}+\rho^i_{22}+\rho^i_{33}+\rho^i_{44}$. All the elements of the density matrixes can be analytically obtained, but for brevity we do not give them in this paper. And the concurrence can be written as
\begin{equation}
\begin{split}
C_i=max\{0,\Omega_i\equiv\frac{2(|\rho^i_{14}|-\sqrt{\rho^i_{22}\rho^i_{33}})}{P_i}\}
\label{eq:one}, (i=1,2,3,4).
\end{split}
\end{equation}
The final success probability of our whole scheme is
\begin{equation}
\begin{split}
P_{fin}=P_1+P_2+P_3+P_4
\label{eq:one},
\end{split}
\end{equation}
and the final concurrence is
\begin{equation}
\begin{split}
C_{fin}=\frac{P_1C_1+P_2C_2+P_3C_3+P_4C_4}{P_{fin}}
\label{eq:one}.
\end{split}
\end{equation}
It is noted that we can still discuss the exact protection and the optimal protection of the entangled state, but for simplicity we only consider the exact protection.
In this paper, we consider the identical decoherence channel $r_a=r_b=0.6$ and the identical pre-weak measurement strength $p^a=p^b$ for both qubits under the exact protection, that is $p_u^a=p_v^a=1-\frac{(1-p^a)(1-r_a)}{p^a}$ and $p_u^b=p_v^b=1-\frac{(1-p^b)(1-r_b)}{p^b}$. It is obvious that each value of $p^a (=p^b)$ uniquely determines a success probability and corresponding concurrence. When we take $p^{i}$ over all the real numbers from $\frac{1-r_{i}}{2-r_{i}}$ to 1, $(i=a, b)$, we obtain $P_{fin}$ (the success probability)-$C_{fin}$ (concurrence) phase diagram.
 It was shown in Ref.~\cite{34} that a scheme based on weak measurement and measurement reversal can also protect an entangled state of two qubits under its exact protecting condition.
 As an example, we compare our result with that of Ref.~\cite{34} in Fig.~8, and for convenience we choose the same entangled state used in Ref.~\cite{34}, i.e., $|\alpha|=0.42$. From Fig.~8 one can see that the success probability of our scheme is much higher than that in Ref.~\cite{34} for the same amount of entanglement, that means the protecting effect of our scheme is much better than that in Ref.~\cite{34}. And the concurrence from our scheme could be approximatively kept at a value of 0.15 with success probability close to 1.

The ESD condition without measurement and feed-forward control is $\sqrt{r_ar_b}\geq|\frac{\alpha}{\beta}|$ for the initial state $\Phi$ with $|\beta|\geq|\alpha|$ and the ESD condition of the scheme in Ref.~\cite{34} is $(1-p^a)(1-p^b)\geq\frac{1}{r_ar_b}|\frac{\alpha}{\beta}|^2$. But in our scheme the ESD never appears.
As decoherence increases, the complete ESD without measurement and feed-forward must occur, and in Ref.~\cite{34} the success probability of preserving some amount of entanglement is approaching zero.
However, our scheme still can preserve part of entanglement with considerable success probability,
in another word, at any degree of decoherence, the ESD never appears for our scheme, which is very different from that of Ref.~\cite{34}.
As an example, we plot the success probability and the entanglement phase diagram for $|\alpha|=0.42$, $r_a=r_b=0.9$ and $p^a=p^b$, and compare our results with that of Ref.~\cite{34} in Fig.~9.
From Fig.~9 one can see that when the concurrence is less than $0.2$, our scheme can obtain a considerable success probability and as the concurrence drops down to 0, the success probability of our scheme will be close to 1, which means that the ESD will never appear in our scheme, but for the scheme of Ref.~\cite{34}, when the concurrence is 0, the success probability is also approaching 0, i.e., the complete ESD occurs in this case.
\begin{figure}
\centering
\vspace{0.5cm}
\includegraphics[scale=0.53]{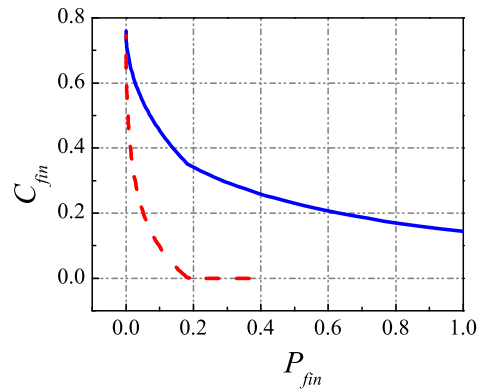}%{enn14.pdf}%{ven12.pdf}%{en11v7.pdf}%{en12120}%{en7.pdf}% Here is how to import EPS art
\vspace{-0.3cm}
\centering
\caption{\label{fig:epsart}(Color online). $P_{fin}$-$C_{fin}$ phase diagram of the effect of protecting entangled state
for $|\alpha|=0.42$, $r_a=r_b=0.6$ and $p^a=p^b$. The blue solid line represents our scheme and the red dashed line represents the scheme in Ref.~\cite{34}. It is noted that in this case the concurrence without any measurement and feed-forward is zero.}
\end{figure}

\begin{figure}
\centering
\vspace{0.6cm}
\includegraphics[scale=0.53]{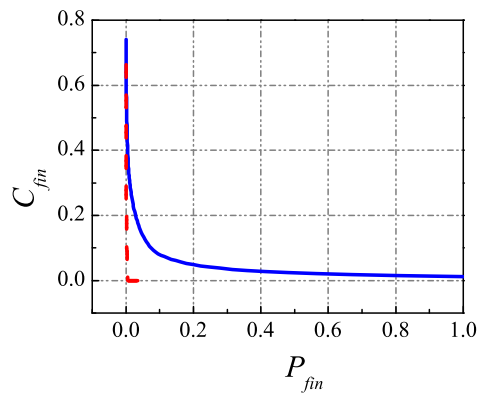}%{enn22.pdf}%{ven23.pdf}%{en21v7.pdf}%{en12122.pdf}%{en8.pdf}% Here is how to import EPS art
\vspace{-0.3cm}
\centering
\caption{\label{fig:epsart}(Color online). $P_{fin}$-$C_{fin}$ phase diagram for $|\alpha|=0.42$, $r_a=r_b=0.9$ and $p^a=p^b$. The blue solid line represents our scheme and the red dashed line represents the scheme in Ref.~\cite{34}. It is noted that in this case the concurrence without any measurement and feed-forward is zero.}
\end{figure}

%%%%%%%%%%%%%%%%%%%%%%%%%%%%%%%%%%%%%%%%%%%%%%%%%%%%%%%%%%%%%%%%%%%%%%%%%%%%%%%%%%%%%%%%%%%%%%%%%%%%%%%%%%%%%%%%%%%%%%%%%%%%%%%%%%%%%%%%%%%%%%%%%%%%
%%%%%%%%%%%%%%%%%%%%%%%%%%%%%%%%%%%%%%%%%%%%%%%%%%%%%%%%%%%%%%%%%%%%%%%%%%%%%%%%%%%%%%%%%%%%%%%%%%%%%%%%%%%%%%%%%%%%%%%%%%%%%%%%%%%%%%%%%%%%%%%%%%%%
\section{EXPERIMENTAL FEASIBILITY}
It should be noted that our scheme can be realized experimentally
with current technology. The experimental realization of the weak
measurement before the noise channel was discussed theoretically in
Ref.~\cite{35}, and it was shown that this kind of weak measurement
can be realized by coupling the system to a meter and performing the
usual projective measurements on the meter only. The experimental
implementation was realized recently in a photonic architecture
\cite{26}. The required weak measurement on the signal qubit
(photon) is realized by entangling it with another meter qubit
(photon), and then a full-strength projective measurement is
performed on the meter qubit, which implements a measurement on the
signal qubit with a strength (ranging from 0 to 1) determined by the
input meter state. The weak measurement after the noise channel has
been demonstrated experimentally \cite{34}. In Ref.~\cite{34}, the
weak measurement (corresponding to weak measurement $\Xi=W_1^+W_1$
after the noise channel in our scheme) and the reversing measurement
(corresponding to weak measurement $\Lambda=N_1^+N_1$ after the
noise channel in our scheme) for a single-photon polarization qubit
are implemented with Brewster-angle glass plates (BPs) and wave
plates. As the BP probabilistically rejects vertical polarization
($|1\rangle_s$) and completely transmits horizontal polarization
($|0\rangle_s$), a single-photon polarization qubit found behind a
BP had been subject to weak measurement or partial collapse
measurement towards $|0\rangle_s$. The reversing measurement is
designed to reverse the effect of weak measurement by making partial
collapse measurement towards $|1\rangle_s$ and it can be implemented
by adding $45^o$ half-wave plates (HWPs) before and after the BPs.
The weak measurement and the reversing measurement strengths can
be varied by changing the number of BPs.

%%%%%%%%%%%%%%%%%%%%%%%%%%%%%%%%%%%%%%%%%%%%%%%%%%%%%%%%%%%%%%%%%%%%%%%%%%%%%%%%%%%%%%%%%%%%%%%%%%%%%%%%%%%%%%%%%%%%%%%%%%%%%%%%%%%%%%%%%%%%%%%%%%%%
%%%%%%%%%%%%%%%%%%%%%%%%%%%%%%%%%%%%%%%%%%%%%%%%%%%%%%%%%%%%%%%%%%%%%%%%%%%%%%%%%%%%%%%%%%%%%%%%%%%%%%%%%%%%%%%%%%%%%%%%%%%%%%%%%%%%%%%%%%%%%%%%%%%%
\section{CONCLUSION AND DISCUSSION}
In this paper, we have proposed a novel scheme of feed-forward control for protecting quantum states against
amplitude-damping noise. We have demonstrated that by using our scheme an unknown quantum state, two known nonorthogonal quantum
states and two-qubit entangled state can be effectively protected. Especially, our scheme is more effective for protecting the
vulnerable qubit states, i.e., the states in which the population in the excited level is higher than that in the ground level,
and works well even for heavy amplitude damping. For some known nonorthogonal states, our scheme can achieve a perfect protection
in an almost deterministic way. And for two-qubit entangled state our scheme can completely avoid the entanglement sudden death no
matter how heavy the amplitude damping noise is.

Now we compare the effect of our scheme with other protecting schemes. In Ref.~\cite{25}, a quantum feedback control scheme
based on weak measurement was proposed to realize the optimal protection of two nonorthogonal states against dephasing noise.
In Ref.~\cite{29}, this method is extended to fighting against other types of noise sources. But this scheme is valid only for
some suitable states and the fidelity has only an improvement less than three percent compared with that without any measurement
and feedback. In contrast, our scheme can protect any quantum state and for some nonorthogonal states can realize almost perfect
protection in an almost deterministic way. Actually our scheme is quite distinctive from that in Ref.~\cite{25,29}, the purpose of
our measurement and feed-forward is totally different from that in Ref.~\cite{25,29}.
Before the noise channel our main purpose for choosing measurement and feed-forward operation is to make the initial state transformed
into the state almost immune to the noise channel and after the channel our aim is to recover the state to the initial state.
In the process, we do not care about the information about the quantum state being protected. But in their works, the objective
to measure is to obtain information about the initial quantum state and then the corresponding feedback operation is to turn the measured state
into the initial state according to the knowledge extracted from the measurement. In other words, our choice of the measurement and
feed-forward only concerns the noise channel, while the measurement and feedback in Ref.~\cite{25,29} mainly concerns the protected
quantum state. On the other hand recently a scheme based on weak measurement and measurement reversal was proposed to protect an
arbitrary qubit state \cite{31} and a two-qubit entangled state \cite{34}. In this paper we have shown that the success probability
for fixed fidelity in Ref.~\cite{31} is lower than ours at any degree of amplitude damping, and our scheme has obvious superiority
over the scheme in Ref.~\cite{31,34} for heavy amplitude damping noise and for the vulnerable states which is easier to deteriorate
in the noise channel. Compared with the result in Ref.~\cite{34} our scheme can achieve higher success probability for keeping the
same amount of entanglement and completely avoid the ESD, while the success probability of preserving some amount of entanglement
for their scheme is approaching zero for heavy damping noise. This can be understood as follows. In our scheme, the pre-measurement
is complete and none of the pre-measurement results has been discarded. In addition, according to any one of the pre-measurement
results, we can not only effectively reserve some information about the initial state which would be used to recover the initial
state in the following steps, but also transform them into these states almost immune to the noise channel.
This characteristics ensures that our scheme can obtain a higher success probability for given fidelity.

It should be pointed out that our scheme can be found applications
in the B92 protocol which is the simplest quantum key distribution
(QKD) protocol conceived by Bennett in 1992 \cite{36}. The B92
protocol is based on only two nonorthogonal states associated with
the two values of the logical bit to be secretly transmitted.
Despite its simplicity, the B92 protocol is considered as a quite
impractical protocol mainly for its low tolerance to noise of a
communication channel. The high dependence on channel noise can be
ascribed to the so-called unambiguous state discrimination)
attack \cite{37}, which represents the principal threat against the
B92 protocol, and severely limits its performances. But one might
use our scheme to overcome this difficulty. Before sending the
signal to Bob, Alice can use our pre-weak measurement and feed-forward
to prepare the state to the state almost immune to the noise, and then
Alice sends it to Bob through the fiber and tells Bob the pre-weak measurement results through classical communication, and when Bob receives the
signal he can use our reversed operation and the reversed post-weak measurement to restore
the information sent by Alice originally. It has been shown that our
scheme can perfectly protect some kind of nonorthogonal states with
almost certainty. In one word, our scheme is very significant for
improving the efficiency and security of quantum communication due
to the fact that the transmission distance is greatly limited by the
serious decoherence of commercial optical fibers now widely used in
quantum communication. It is worth noting that our scheme is
entirely feasible with current technology. In the end, we also hope
that the idea of our feed-forward control scheme in this paper provides new thinking for fully utilization of quantum measurement-based control method in the future.\\

\begin{center}
\textbf{ACKNOWLEDGMENT}
\end{center}
\par This work was supported by the National Natural Science
Foundation of China (Grants Nos. 11274043, 11375025).

%%%%%%%%%%%%%%%%%%%%%%%%%%%%%%%%%%%%%%%%%%%%%%%%%%%%%%%%%%%%%%%%%%%%%%%%%%%%%%%%%%%%%%%%%%%%%%%%%%%%%%%%%%%%%%%%%%%%%%%%%%%%%%%%%%%%%%%%%%%%%%%%%%
%%%%%%%%%%%%%%%%%%%%%%%%%%%%%%%%%%%%%%%%%%%%%%%%%%%%%%%%%%%%%%%%%%%%%%%%%%%%%%%%%%%%%%%%%%%%%%%%%%%%%%%%%%%%%%%%%%%%%%%%%%%%%%%%%%%%%%%%%%%%%%%%%%

%\newpage %Just because of unusual number of tables stacked at end

\end{document}